# Bunch-by-bunch Beam Transverse Feedback Electronics Designed for SSRF

Jinxin Liu, Lei Zhao, *Member, IEEE*, Linsong Zhan, Shubin Liu, *Member, IEEE*, and Qi An, *Member, IEEE*

***Abstract*—Shanghai Synchrotron Radiation Facility (SSRF), is one of the third-generation high-beam current (3.5 GeV) synchrotron light sources. In the storage ring of SSRF, multi-bunch instabilities would increase beam emittance and energy spread, which degrade beam quality and even cause beam loss. To address the above issues, a Transverse Feedback System is indispensable for SSRF, in which the key component is the bunch-by-bunch transverse feedback electronics. The whole feedback system consists of five main parts: Beam Position Monitor (BPM), RF front-end, signal processor, RF amplifier, and vertical/horizontal transverse kickers. The dissertation focuses on the signal processor we design, which is the main part of the feedback electronics. We conducted initial testing on the signal processor to evaluate its performance and function. Test results indicate that ENOB of the Analog-to-Digital Conversion circuit is better than 10 bit with 100 MHz input signal, and remains better than 9.5 bit up to 300 MHz, which is good enough for the required 7.9 bit; the frequency response of the whole system also concords well with the simulation results, and the suppression in amplitude response at the critical frequency points is better 35 dB while the uncertainty of phase response is better than 2º, all meeting the application requirement.**

***Index Terms*—SSRF, transverse feedback, bunch-by-bunch feedback**

## I. Introduction

SHANGHAI Synchrotron Radiation Facility (SSRF) is one of the third-generation high-beam current (3.5 GeV) synchrotron light sources. In the storage ring of SSRF, multi-bunch instabilities would increase beam emittance and energy spread, which degrade beam quality and even cause beam loss. To address the above issues, a Transverse Feedback System is indispensable for SSRF [1].

A total of 720 bunches circulates in the tunnel with a duty ratio of 500:220 and a Turn-By-Turn (TBT) frequency of $f_{mc}$=693.964 kHz (this frequency is synchronized to the system machine clock). This corresponds to a bunch-by-bunch frequency of 499.654 MHz (720×$f_{mc}$). There exists transverse position oscillation for each bunch, with the same frequency but different phase & amplitudes. Therefore, a feedback system is required to process the beam signal of each bunch and generates corresponding feedback signals. The key parameters for the feedback system are shown in Table I [2-3]. The oscillation frequencies in vertical and horizontal position are 11.29 and 22.22 normalized by $f_{mc}$.

TABLE I
KEY FEEDBACK PARAMETERS OF THE SSRF

| Key Parameters | Quantity |
|---|---|
| RF frequency | 499.654 MHz |
| Harmonic number | 720 |
| Vertical tune | 11.29 |
| Horizontal tune | 22.22 |

The whole feedback system consists of five main parts: Beam Position Monitor (BPM), RF front-end, signal processor, RF amplifier, and vertical/horizontal transverse kickers, as shown in Fig. 1 [4].

The RF front-end imports the signals from the BPM, and then filters them. The signals are input to the signal processor to calculate the feedback coefficients that are converted to controlling voltages using high-speed Digital-to-Analog Converters (DACs) [5]. These voltages are then amplified and used as the input of the kickers to tune the beam into the optimum orbit. The main part in the feedback electronics is the signal processor, which is discussed in the following sections [6].

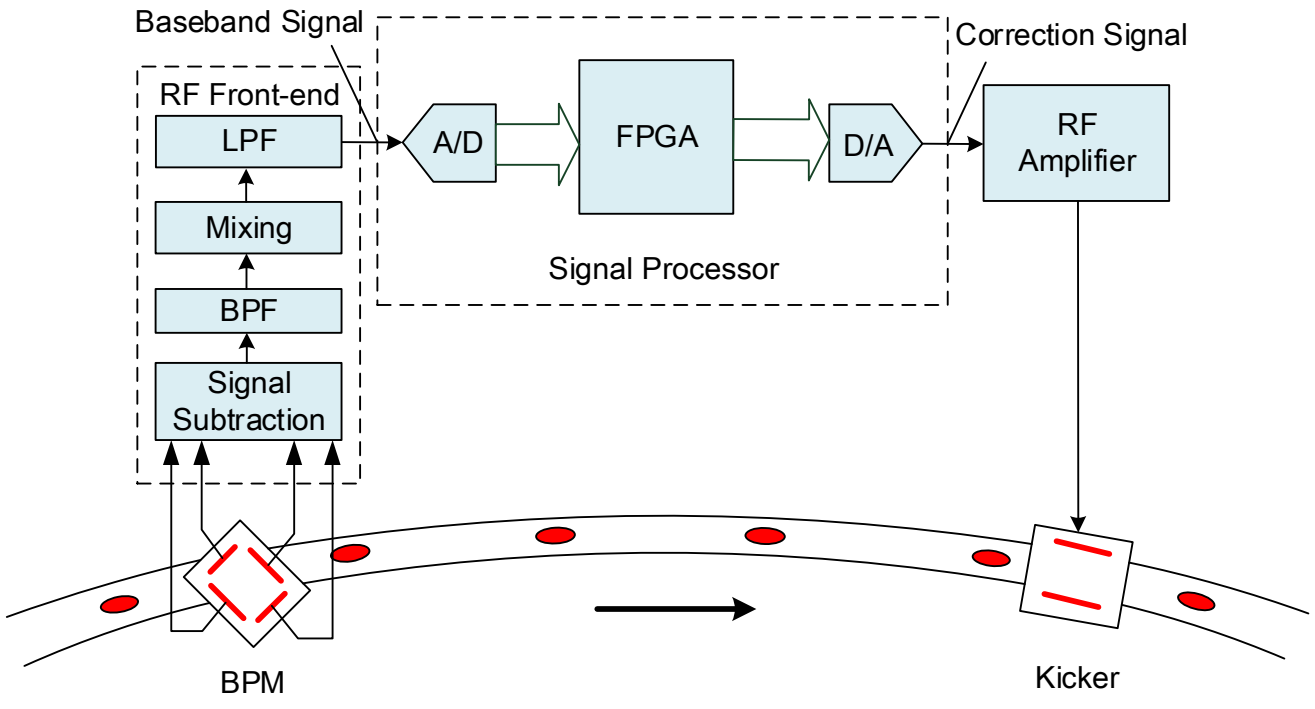

Fig. 1. The overview over the transverse feedback system.

Manuscript received June 24, 2016; revised March 26, 2017.

This work was supported by Knowledge Innovation Program of the Chinese Academy of Sciences (KJCX2-YW-N27), National Natural Science Foundation of China (11205153) and CAS Center for Excellence in Particle Physics (CCEPP).

The authors are with the State Key Laboratory of Particle Detection and Electronics, University of Science and Technology of China, Hefei, 230026; and Modern Physics Department, University of Science and Technology of China, Hefei, 230026, China (telephone: 086-0551-63607746, corresponding author: Lei Zhao, e-mail: zlei@ustc.edu.cn).

## II. System Design

### *A. Structure of the Signal Processor*

Fig. 2 shows the structure of the signal processor.

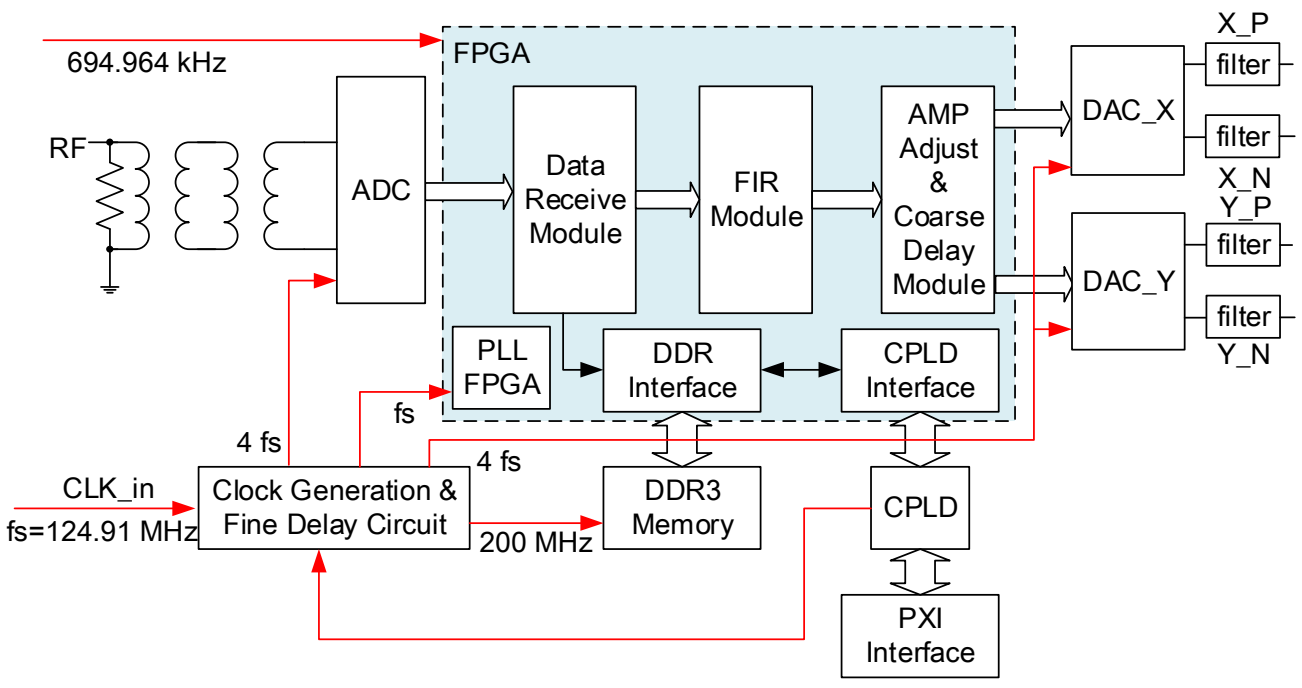

Fig. 2. Block diagram of the signal processor.

The signals from the BPM are pulses with a repetition frequency of 499.654 MHz, and each pulse corresponds to one beam bunch. The sampling frequency of the Analog-to-Digital Converter (ADC) is also set to 499.654 MHz to digitize the bunch-by-bunch signal. PLL based clock generation circuits are designed to synchronize the sampling clock with the system of the accelerator, as shown in Fig. 2. The input signals of the signal processor are the outputs of the RF front-end which filters the signals from BPM to a bandwidth below 250 MHz. A 12 bit 500 Msps ADC chip AD9434 is employed in this design. The output of the ADC are then fed to a Field Programmable Gate Array (FPGA) to calculate the feedback coefficients based on Finite Impulse Response (FIR) filters in [7-10]. These coefficients are used as the input of two 14 bit DAC (AD9736 from Analog Device Inc.) working at 500 MHz, which are used for transverse feedback in the vertical and horizontal (i.e. X and Y in Fig. 2) position, respectively.

### *B. Clock generation and delay adjustment Circuits*

The feedback system requires for a strict timing [11]. First, the ADC should sample at approximately the peak of the analog signal generated when a bunch crosses the BPM. A/D Conversion and the following digital signal processing in Fig. 3, as well as DAC would all contribute to the latency between the input signal and the output signal of the processor. To make sure that the kicker functions at the correct bunch, we must also precisely adjust time delay of the processor output signal. To achieve this, we designed the clock generation and delay adjustment circuits, as shown in Fig. 3.

As shown in Fig. 3, this processor receives the 125 MHz system clock from the accelerator, and employs one Phase Locked Loop (PLL) to generate a 500 MHz ($f_s$) sampling clock for the ADC. To precisely adjust the sampling clock phase, a delay line chip (Delay1 in Fig. 3, SY89295 with 10 ps step size) is used [12]. Since this adjustment is within one period of the 2 ns sampling clock, the 11.6 ns range of SY89295 is large enough in this design.

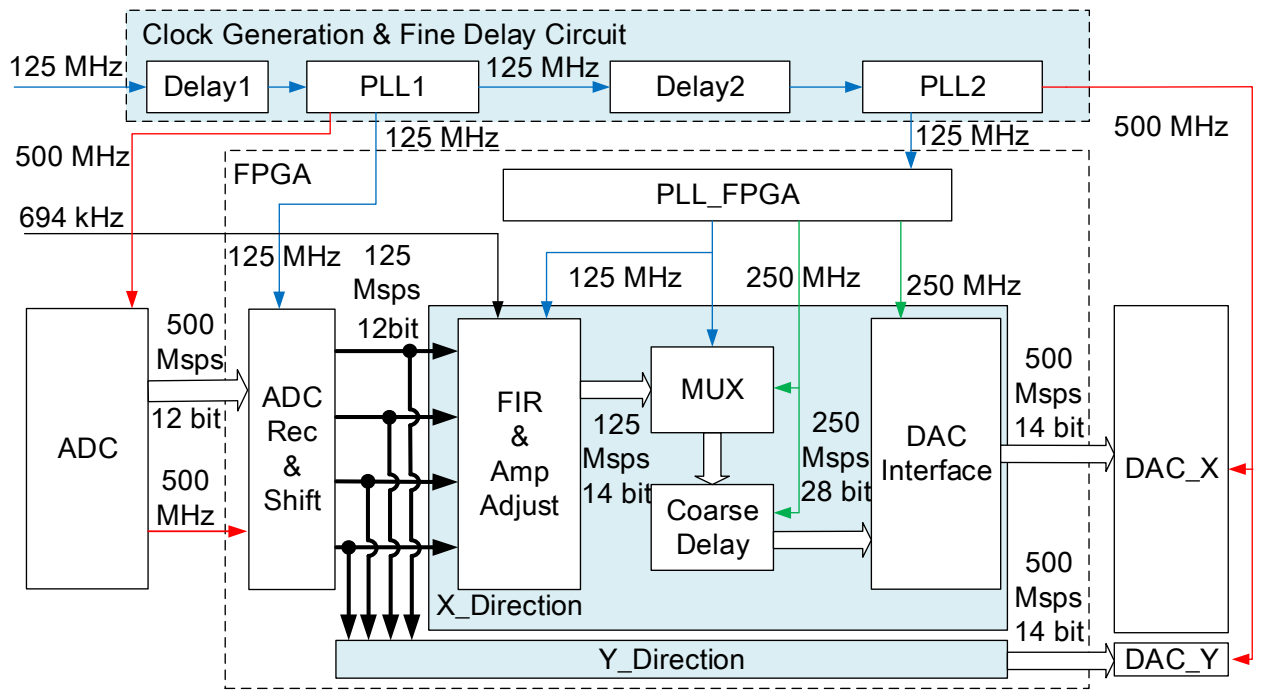

Fig. 3. Block diagram of the clock generation and delay adjustment circuits.

As for the delay adjustment for the processor output signal, a dynamic range up to 1 440 ns ($720\times1/f_s$) is required. We achieve this in two stages. The first stage is coarse time adjustment (marked as "Coarse Delay" in Fig. 3) between the MUX block and the DAC interface block. The 125 Msps data streams from "FIR & Amp Adjust" block are converted to 250 Msps by the Multiplexer (MUX in Fig. 3), delayed and then fed to the DAC interface. The coarse delay is implemented with a 9 bit counter working at 250 MHz, and an adjustment range of 2 μs & a step size of 4 ns can be achieved [13-14]. As for the fine time adjustment, another delay line chip is placed between PLL1 and PLL2, which is able to achieve 10 ps delay adjustment precision. Besides the 125 MHz clock signals for FPGA, PLL2 is also responsible for jitter cleaning and generating high quality 500 MHz clock signals for the DACs. Finally a precise adjustment with a step size of 10 ps can be accomplished both for the sampling of bunch signals and the time of feedback signal arriving at the kicker. Since all the delay circuits are all integrated in the electronics and can be configured according to FPGA commands, better system flexibility is achieved.

This processor also receives a synchronous 694.964 kHz clock, and uses it as the reference to identify each bunch position in the data stream for specific processing on single bunch. This will be discussed in detail in the following sub section.

### *C. Digital Signal Processing Algorithm*

The block diagram of the Digital Signal Processing (DSP) algorithms in the FPGA is shown in Fig. 4. The data from the ADC are in a 500 Msps stream, in which each sampling point corresponds to the information of a certain bunch. This stream is first deserialized to four 125 Msps data streams [15], and then the information of each bunch is extracted from these data streams using shift registers, and is fed to FIR filter to calculate the corresponding feedback coefficient. This coefficient can also be adjusted bunch by bunch with different gains for machine research.

The information of each bunch in the original 500 Msps ADC output data stream is required to be separated out to be processed. Considering the logic resource consumption, it is not possible to implement an individual FIR filter for each bunch in the FPGA device. Meanwhile, the original 500 Msps data should be converted to data streams at a lower data rate, considering the reliable work frequency of the logic in FPGA.

Therefore, our solution is to deserialize the 500 Msps data to four 125 Msps streams, and employ a parallel structure FIR filter to process each data stream bunch by bunch in a time-sharing mode.

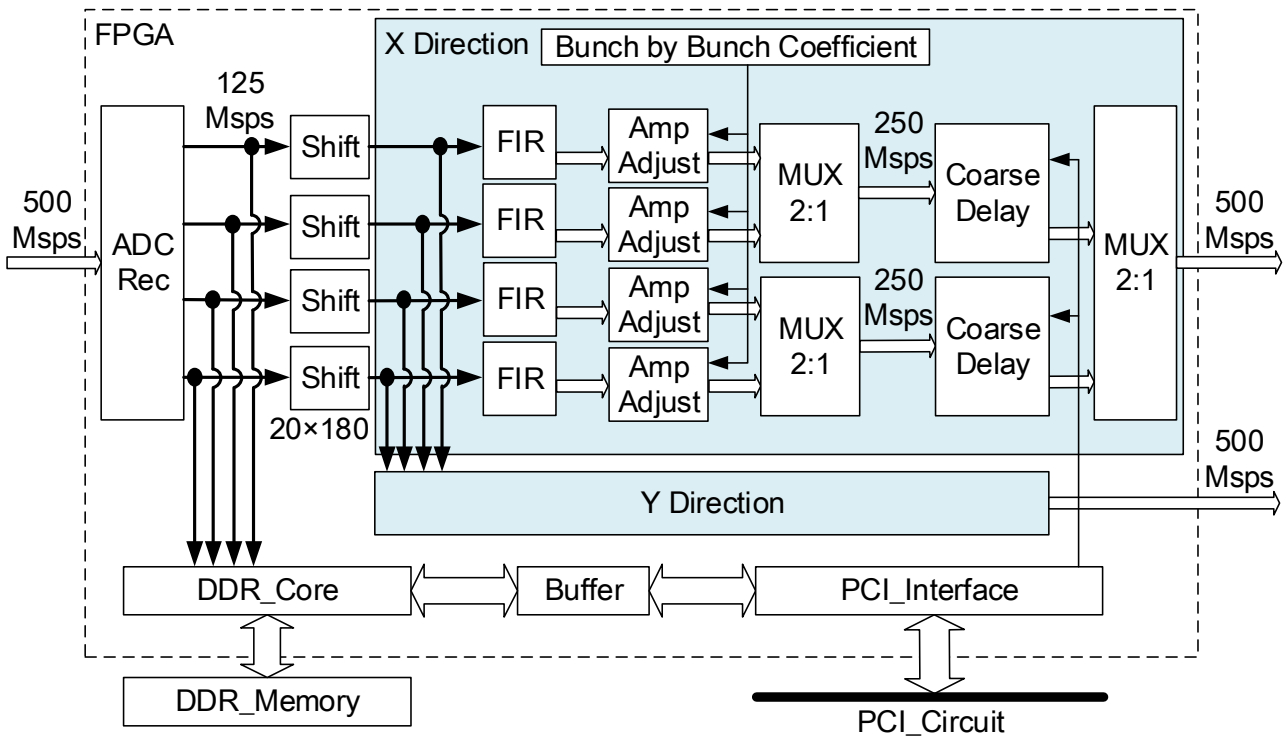

Fig. 4. Block diagram of the digital signal processing algorithm.

The information of each bunch in the original 500 Msps ADC output data stream is required to be separated out to be processed. Considering the logic resource consumption, it is not possible to implement an individual FIR filter for each bunch in the FPGA device. Meanwhile, the original 500 Msps data should be converted to data streams at a lower data rate, considering the reliable work frequency of the logic in FPGA. Therefore, our solution is to deserialize the 500 Msps data to four 125 Msps streams, and employ a parallel structure FIR filter to process each data stream bunch by bunch in a time-sharing mode.

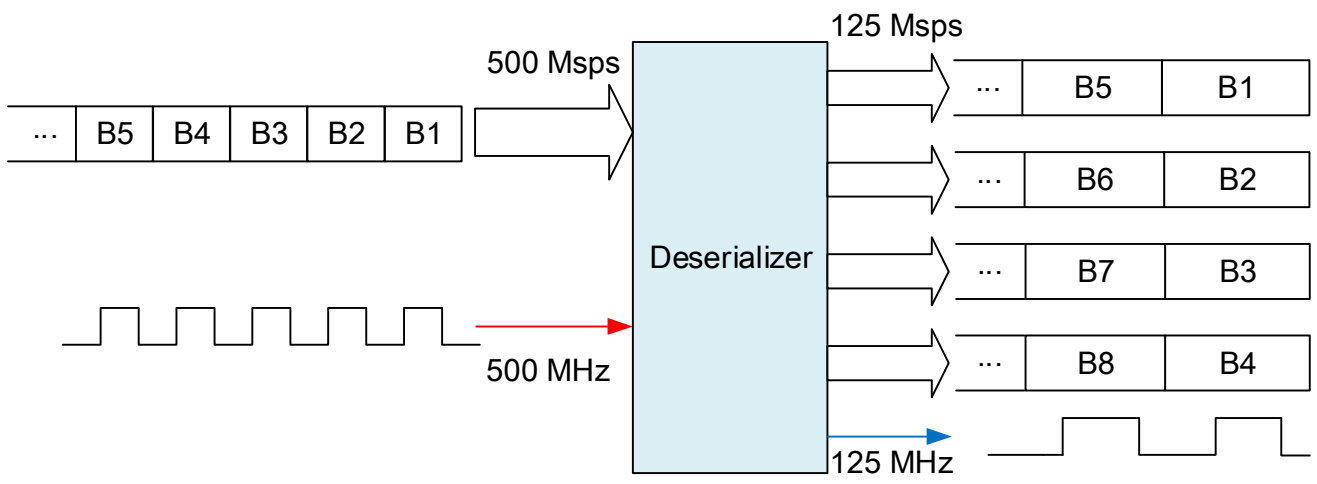

Fig. 5. Block diagram of the deserialization

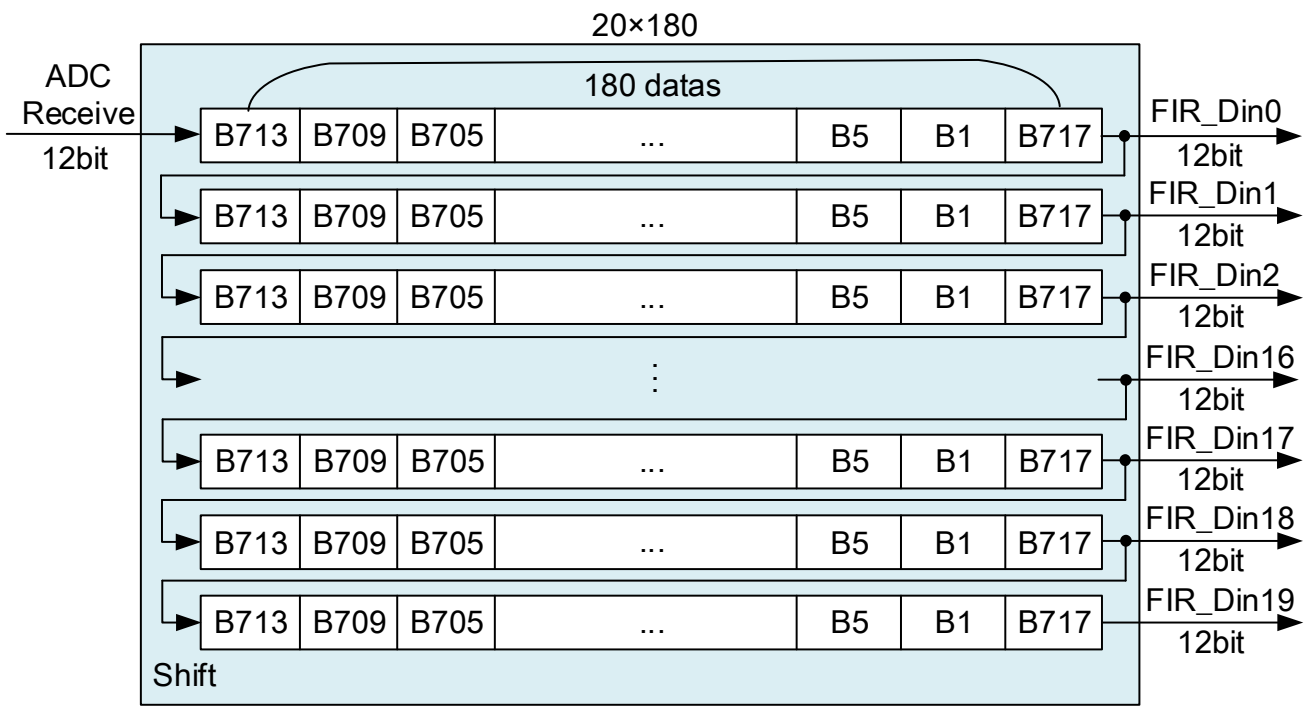

Fig. 6. Block diagram of the shift register array.

After deserialization, the position of the data for each bunch (marked as "B1", B2", ….) is changed to the pattern illustrated in Fig. 5. In this design, 20-order customized double-zeros FIR filters are required [16]. The FIR filters require that the 20 inputs should be the data of each bunch in 20 continuous cycles. Correspondingly, we need to regroup the 125 Msps data stream to meet this requirement. As shown in Fig. 6, we designed 20 shift registers, each with a depth of 180 (corresponding to bunch number within one turn, i.e. 720/4), and connected them together head to tail. With this design, the data for each bunch ("B717" for example in Fig. 6) will appear in the 20 parallel output ports, and bunch position will shift one by one to the output ports, which are connected to the FIR filter inputs.

The key of the algorithm is the feedback coefficient calculation. A family of 20-tap FIR filters are designed to accomplish the calculation.

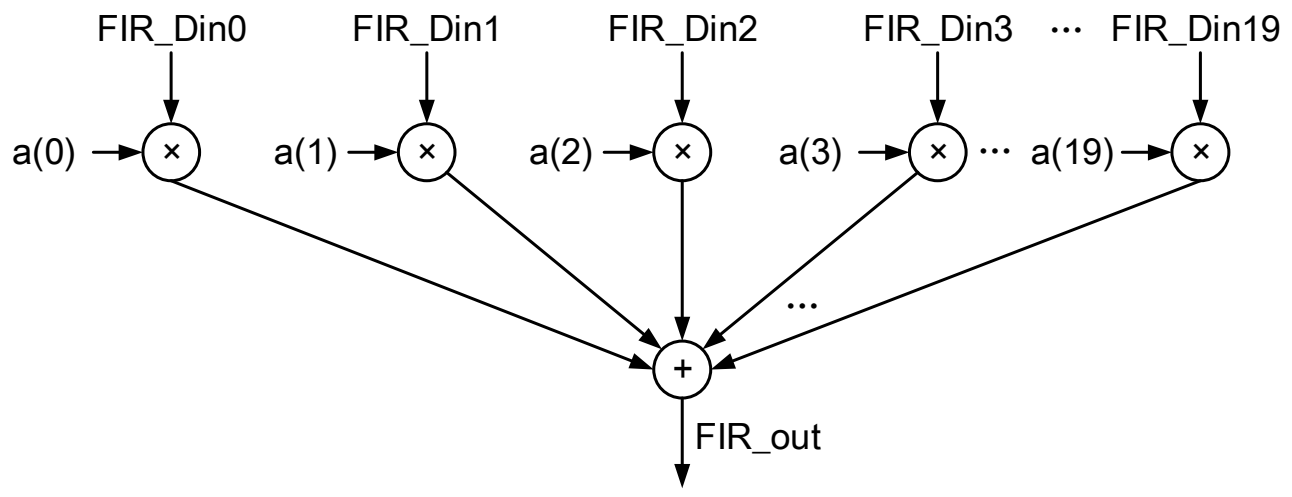

Fig. 7. Structure of the FIR filter.

The structure of the FIR filter is shown in Fig. 7. The 20 output data from the shift register array are multiplied directly with different FIR parameters (a(0), a(1)…a(19)), and then summed together to obtain the coefficient for the bunch. Since it does not require delay elements in common FIR filters (e.g. the FIR core in Xilinx FPGA [17]), the coefficient calculation can be finished within one clock period, and thus multiple bunches can be processed using only one filter at interleaved time point.

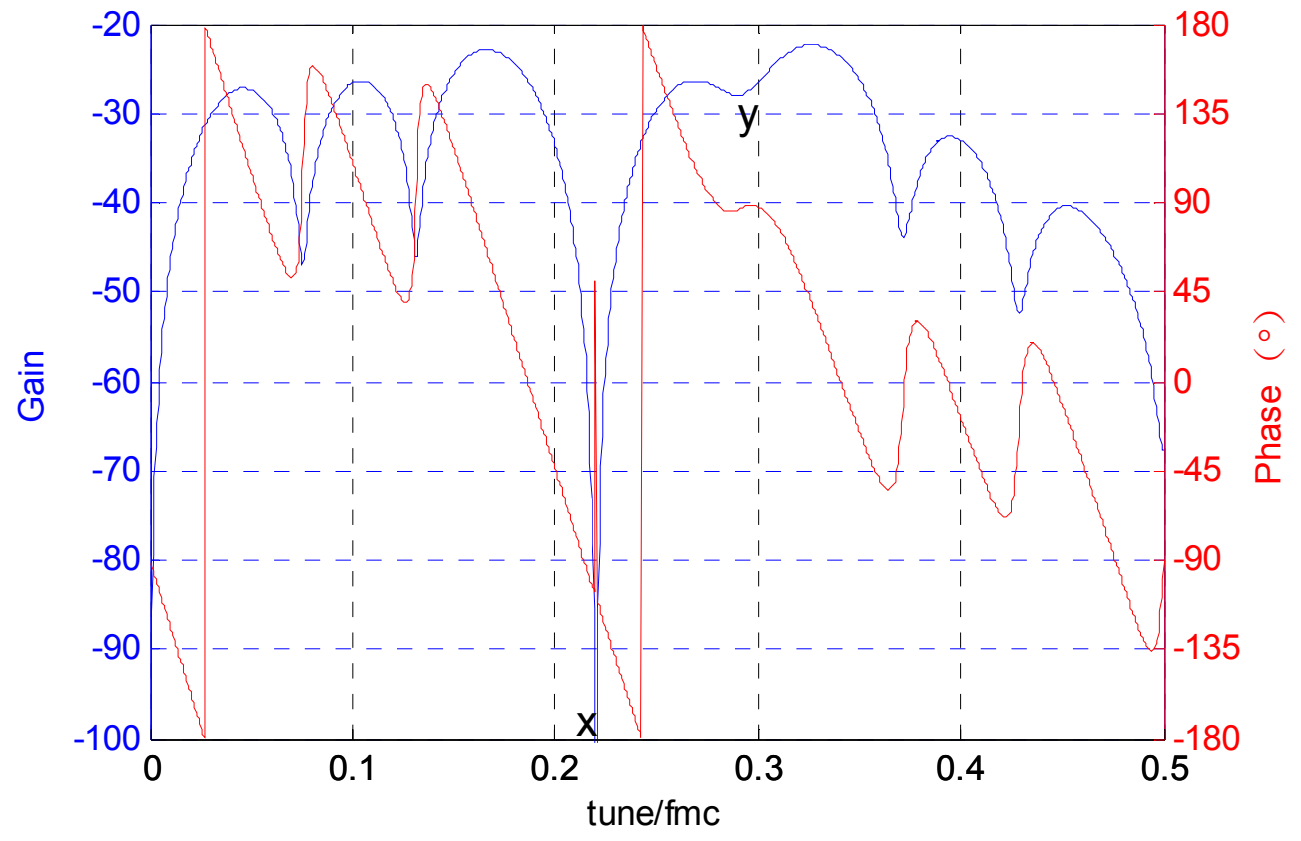

Fig. 8. The spectrum of simulating frequency response.

After the design of the FIR filter, we conducted simulations to verify its function. Fig. 8 shows the amplitude and phase frequency response of the 20 tap FIR filter [18] for vertical direction (Y-direction). The label "tune" in X axis of Fig. 8 refers to the normalized frequency of the FIR response (for example, tune=0.5 means $\omega = \pi$ ). Since the inputs of FIR filter are the data of each bunch in 20 continuous cycles, this "tune" value is also the normalized frequency of FIR sampling frequency $f_{mc}$ (for example, tune=0.5 means $f=f_{mc}/2$). Since the sampling frequency is less than half of the oscillation frequency ($22.22\times f_{mc}$ in X direction and $11.29\times f_{mc}$ in Y direction of beam transverse position), spectrum shifting occurs (i.e. high frequency signals are moved to low frequency locations). After processing of the FIR filters, the positon oscillation frequency is moved from 11.29 to 0.29 in Y direction (i.e. vertical direction) and from 22.22 to 0.22 in X direction (i.e. the oscillation frequencies in X and Y direction correspond to the points at tune=0.22 and 0.29 in Fig. 8).

To get the X-direction and the Y-direction feedback coefficient simultaneously and separately, the amplitude-response in X direction should be zero at vertical tune point (0.29), and vice versa [19-20]. Another zero point is at DC in Fig. 8 where the close orbit component resides. This zero point means that the irrelevant energy is removed, and thus the SNR (Signal to Noise Ratio) of the feedback signal can be improved. This FIR structure is named double-zeros filter, and the simulation results concord well with the expected. Another feature is that the phase response should be $\pi/2$ at the nominal tune (i.e. 0.22 in X direction) to get the best feedback effect [21]. Of course, this phase shift value also includes the contribution caused by hardware delay. To address the above issue, the FIR parameters can be modified via remote PC, according to the calibration results. Besides, both the amplitude and phase response should be almost static within a small region as the margin for the possible slight uncertainty of the nominal tune point. The simulation results concord with the expected.

During machine research, it is expected that the amplitude of each bunch can be adjusted individually with different gain factors. And on-line configuration of the gain values is preferable. To achieve this, we designed a circular memory, as shown in Fig. 9. This memory is implemented based on a RAM structure, and each cell in it contains the gain factor for each corresponding bunch. A read pointer circulates along these 720 cells, and the gain factor is read out and applied to the multiplication block ("Amp Adjust" in Fig. 4) bunch by bunch. Since it is a RAM, the content can be modified on line by users if needed. To make sure the read pointer is in the correct position, bunch identification should be implemented in the data stream. This is accomplished using the 693.964 kHz machine clock, as mentioned above.

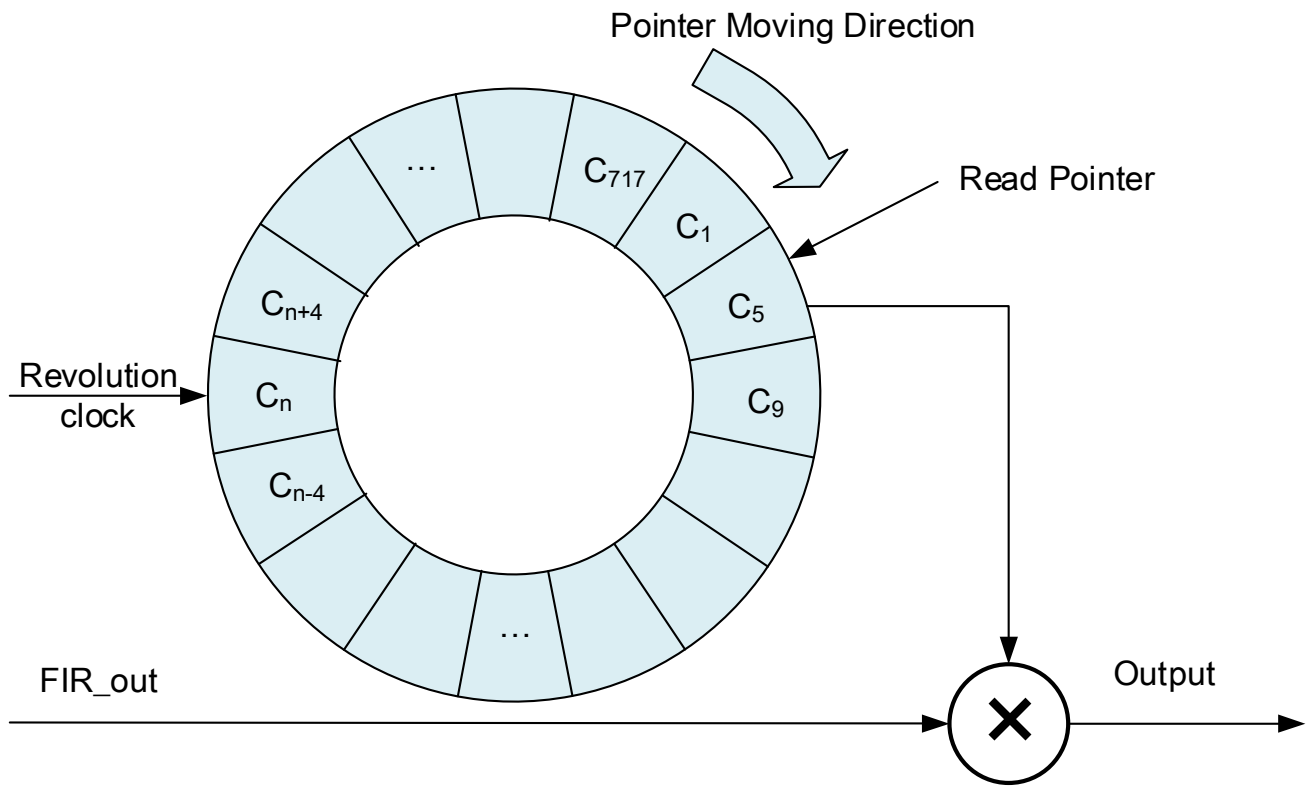

Fig. 9. Block diagram of the amplitude adjust block.

All the digital signal processing algorithm and data interface are integrated in a Xilinx FPGA XC6VLX365T_FF1156. As for the PXI interface, an Altera CPLD EPM2210F324C5 is applied to accomplish this.

Fig. 10 shows the photo of the feedback electronics.

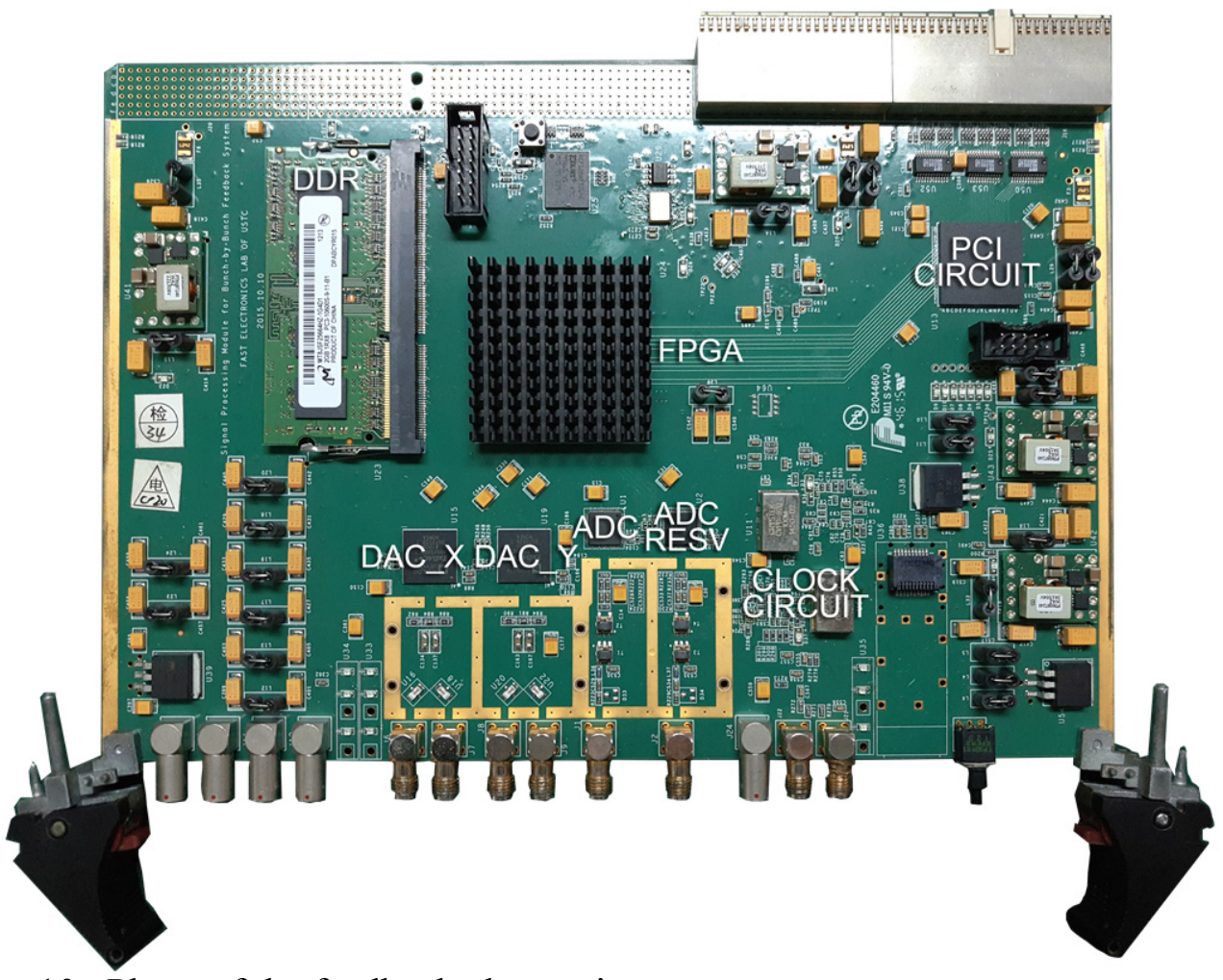

Fig. 10. Photo of the feedback electronics.

## III. System Performance Test

### A. *Test system setup*

We also conducted initial testing on the signal processor to evaluate its performance and function.

Fig. 11 shows the system under test in the laboratory. The electronics is designed as a standard PXI-6U module. The feedback electronics can be controlled by a remote PC through Ethernet via the Single Board Computer (SBC) in Slot 0 of the PCI crate.

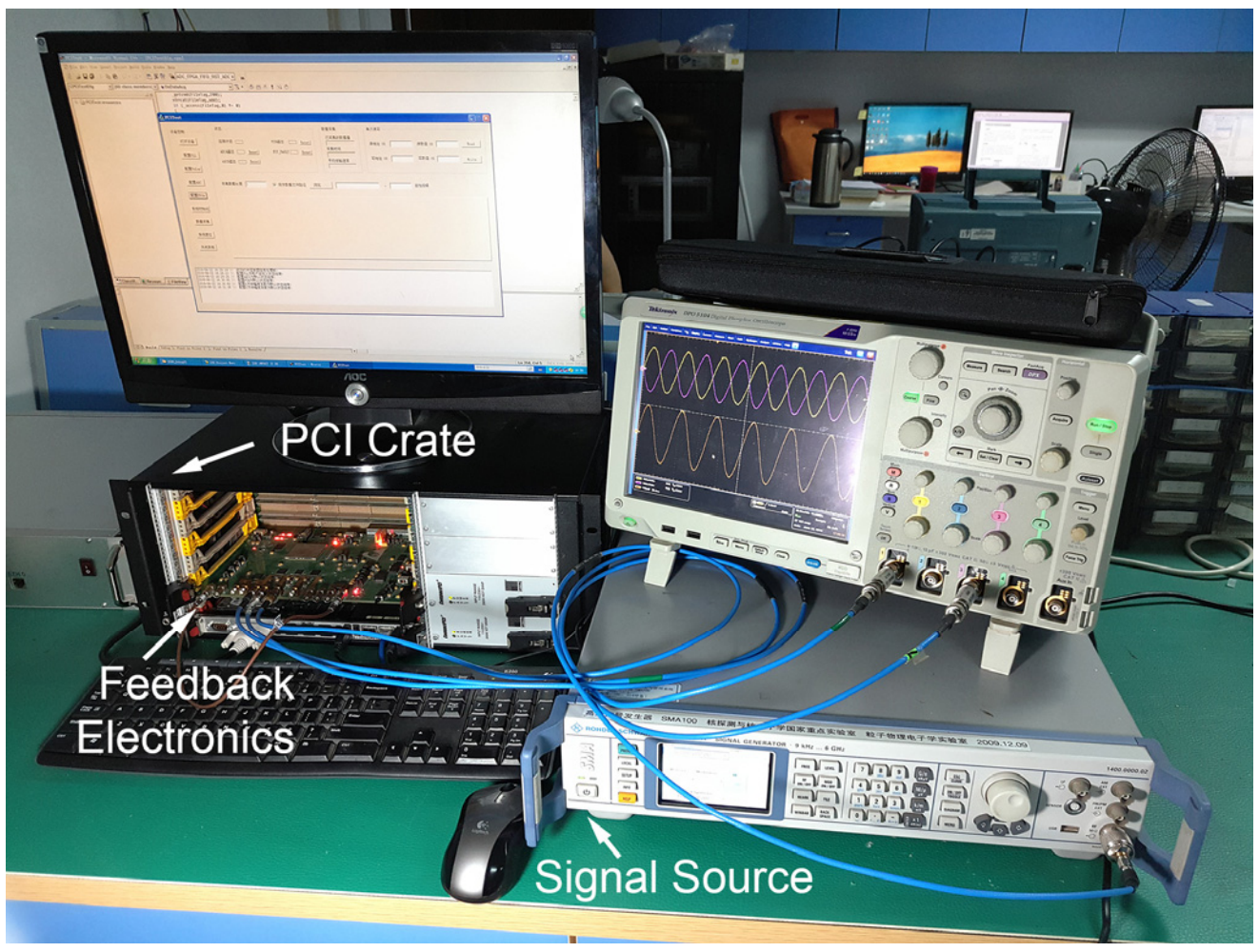

Fig. 11. System under test.

### B. *Performance Test Results*

#### 1) *ADC Performance Test*

The high-speed high-resolution A/D conversion is a crucial part in the system. We performed the ADC dynamic analysis based on the IEEE Std.1241-2010 [22] to assess the circuit performance. SMA 100A, a high performance signal source, is used to generate input sine wave signals from 10 MHz to 800 MHz. Then the signal is filtered by a coaxial BPF (Band Pass Filter) and imported to the signal processor for digitization.

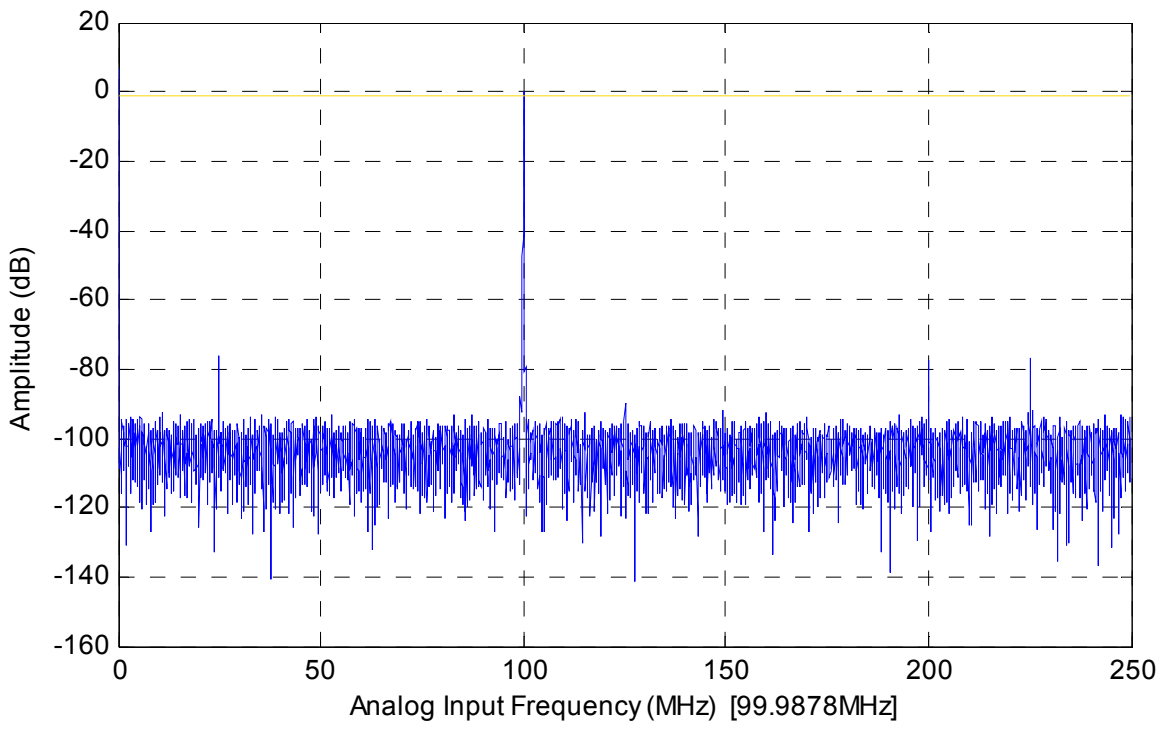

Fig. 12. Typical frequency spectrums of the ADC output.

Fig. 12 shows the typical frequency spectrums (with input frequency of around 100 MHz) of the ADC output signals. The SINAD (Signal-to-Noise and Distortion Ratio) is 63.69 dB and the ENOB (Effective Number of Bit) is 10.3 bit.

To make sure that the electronics is sensitive enough to detect and react to the oscillation of each bunch, the A/D conversion circuits are required to distinguish residual oscillation value (about 40μm) of the beam position. There exists an equation between the detecting resolution and the SINAD of the circuits, as in

$$dx = k/(SINAD \times 2) \quad (1)$$

where dx is the detecting resolution, and k is 11.6 mm in the system. It is expected that dx can be as low as 20 μm. From (1), it can be calculated that the SINAD is required to be better than 49.2 dB, i.e. an ENOB of 7.9 bit. We prefer dx can be as low as 10 μm, and the ENOB is 8.9 bit after calculation.

Considering that analog bandwidth requirement of RF front end is 250 MHz, we also changed the input signal frequency and conducted a series of test results on A/D conversion performance, as shown in Fig. 13. The results indicate that an ENOB better than 9.5 bit in the input frequency range up to 300 MHz is successfully achieved, which is good enough for the required 7.9 bit.

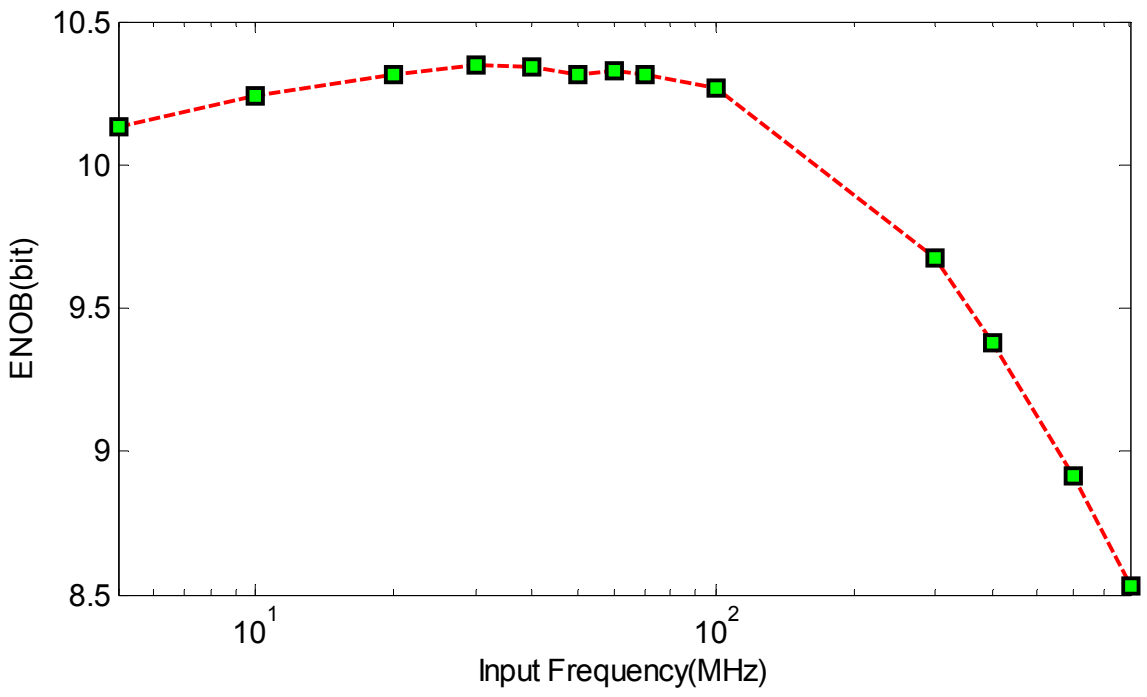

Fig. 13. ENOB curve along with input frequency changing.

*2) DAC Performance Test*

Performance of the D/A conversion circuit also directly influences quality of the feedback signal. Fig. 14 shows the DAC output voltages with different input codes. A good linearity is observed according to the test results, and the Integral Non-Linearity (INL) is better than 0.165%FS.

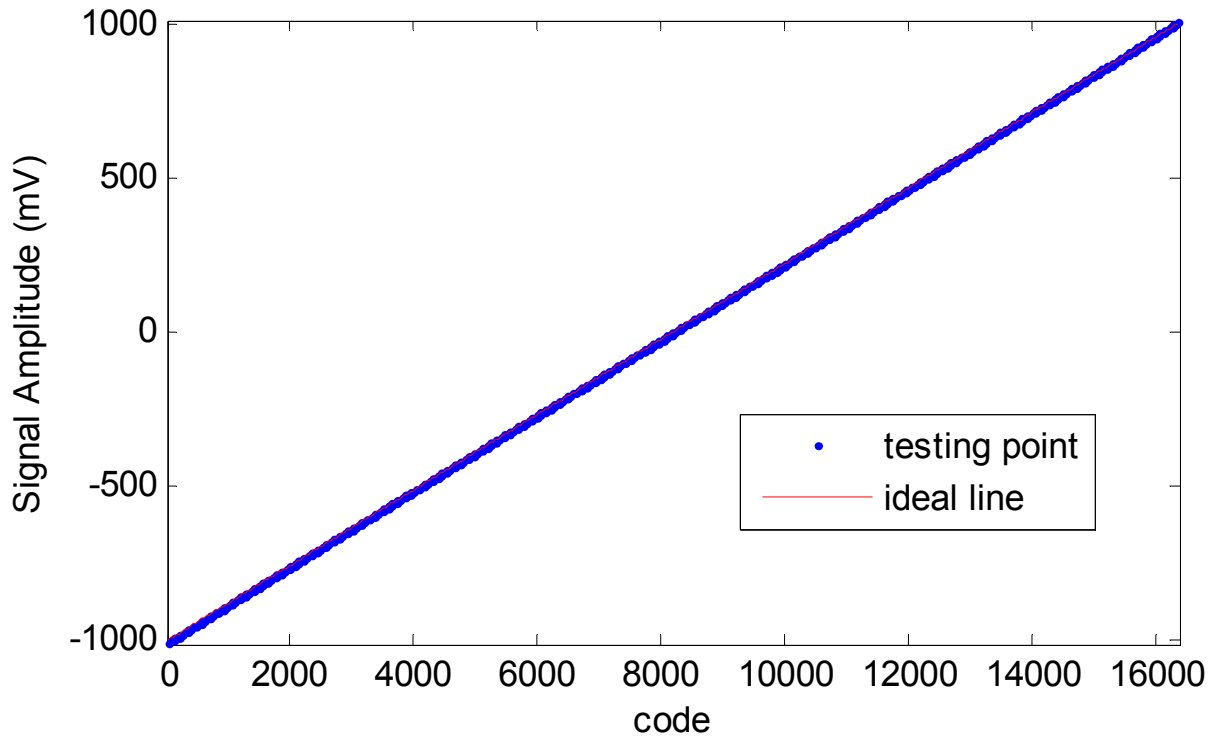

Fig. 14. The DAC output voltages with different input codes.

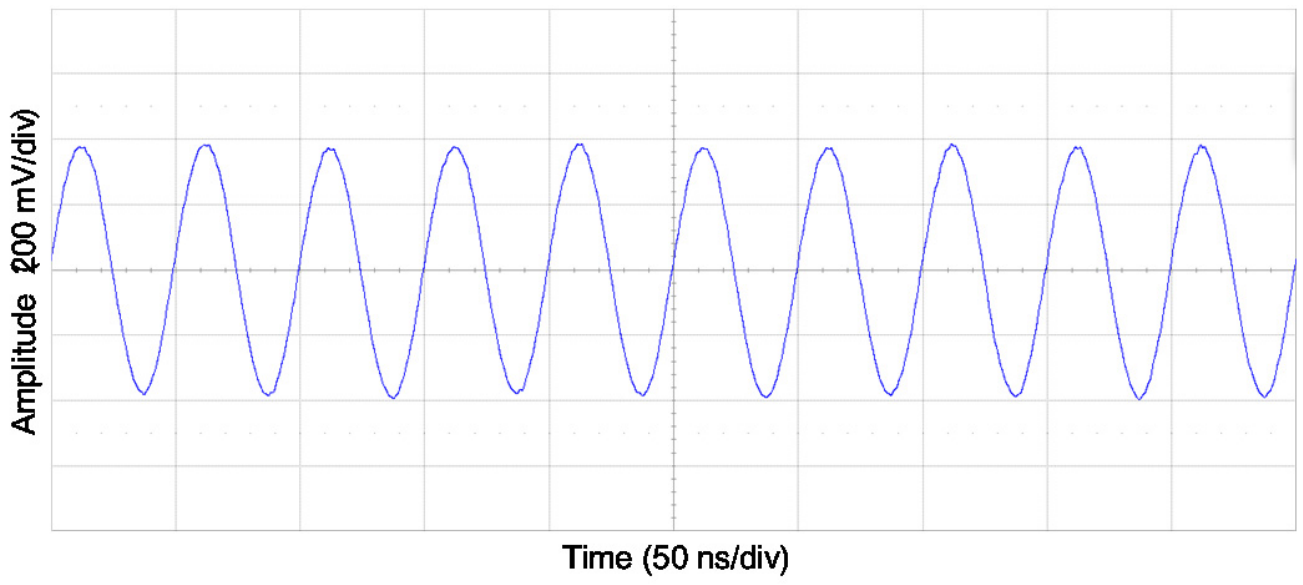

Fig. 15. Typical DAC output waveform in time domain.

We also conducted tests with DAC output sinewave signal. In this test, we imported a 20 MHz sinusoid signal to the ADC of the processor, this signal is digitized and then fed to DAC as the input codes. The DAC output waveform (after a 250 MHz Low Pass Filter) is shown in Fig. 15.

*3) Time Delay Test*

Since precise timing is very important for the transvers feedback system, tests were also conducted to evaluate the delay adjust ability of the delay line chip, which is the key part for this function.

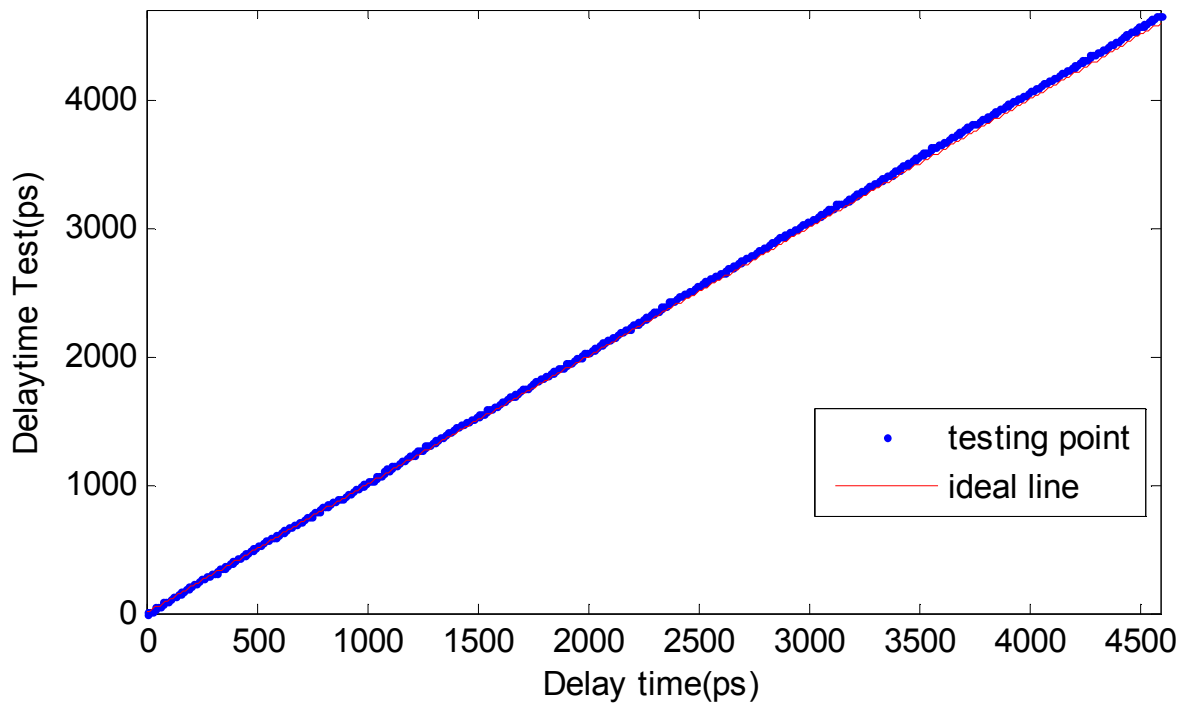

Fig. 16. Delay test results with different input codes.

In the test, DPO 5104, a high sample rate high bandwidth (sample rate: 10 G/s, bandwidth: 1 GHz) oscilloscope is used to measure the delay between the input signal and output signal of delay line with the delay code changing.

Fig. 16 is delay test results with different input codes, and the delay adjustment step size is within 10 ps, which is good enough for the application.

### 4) *Functionality Test*

We also conducted tests to evaluate the performance of the whole feedback signal processor.

We established a model according to the requirement of the signal processor on the platform of MATLAB, and then we obtained the frequency response of the expected feedback function. Then we used the network analyzer Agilent E5071C to obtain the actual frequency response of the signal processor by sweeping frequency in the range from 0 to 356.982 kHz. This frequency corresponds to the turn-by-turn repetition frequency of 693.964 kHz (499.654 MHz/720), since 720 bunches are circulated in the storage ring with a duty ratio of 500:220.

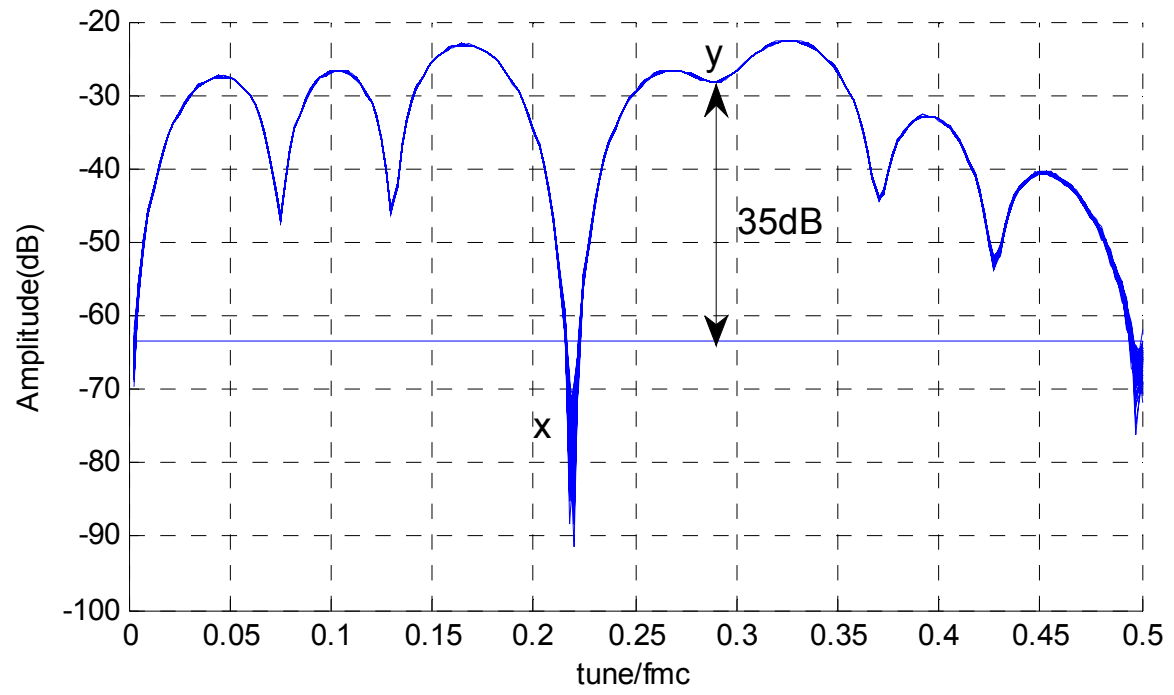

(a)

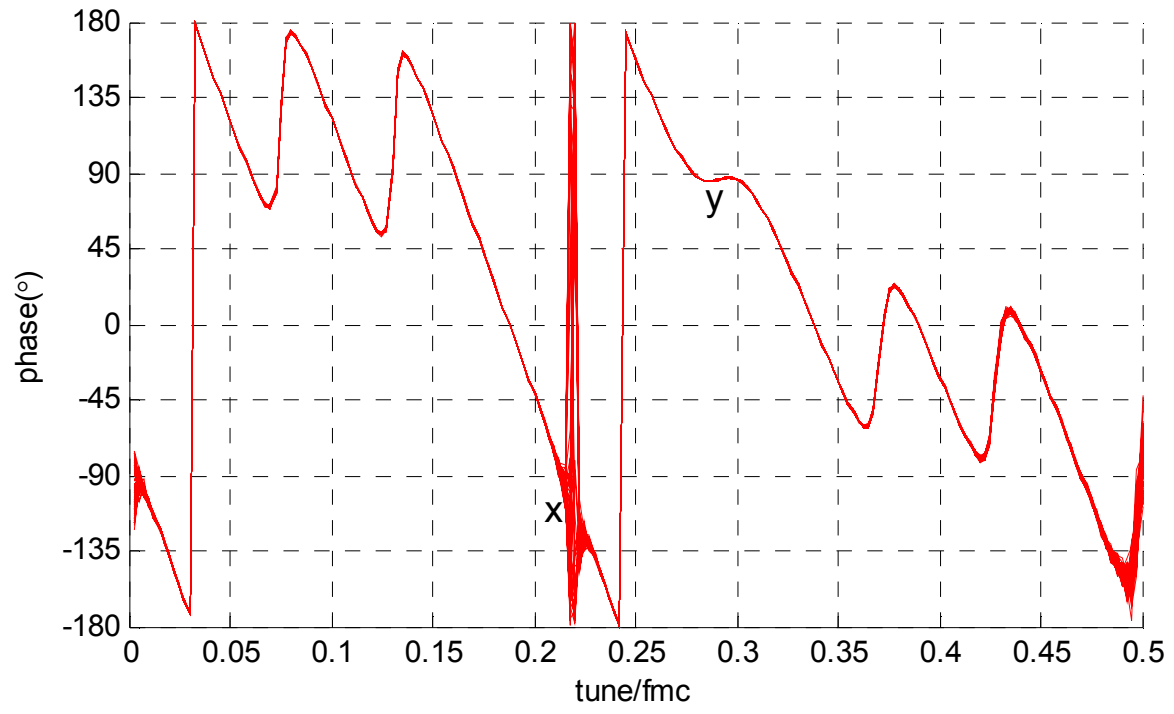

(b)

Fig. 17. The spectrum of be as low as 20 μm frequency response test results. (a) and (b) is the amplitude-frequency response and phase-frequency response, respectively

Fig. 17 shows the test results of representative vertical frequency response. The blue curve is the amplitude-frequency response, and the red curve is the phase-frequency response. The results concord well with Fig. 8.

We conducted tests for 100 times and plot the results together. It can be observed that the suppressions at tune=0.22 compared with tune=0.29 are all better than required 35 dB for these 100 times of test results.

As for the phase-frequency response, the response should be close to 90º at tune=0.29, which is optimal for beam feedback. The actual phase response would deviate from 90º [23], which would be caused by the time period when the bunch passes the distance between the BPM pickup and kicker. To adjust the phase response back to the correct value, we designed 32 sets of FIR coefficients, which correspond to 32 possible phase values with an interval of 11.25º (i.e. 360º/32). With this design, the worst case phase error is 11.25º/2, meaning a feedback error of 0.5% (i.e. sin(90º)-sin(90º±11.25º/2)[24]). As for Fig. 17 (b), the requirement for the system is that the phase-frequency response at tune=0.29 should be stable within ±11.25º/2. The test results indicate that the uncertainty of the 100 times of test results is within 2º, so it also meets the requirement.

## IV. Conclusion

A bunch-by-bunch beam transverse feedback electronics was designed for the SSRF. Through high-speed high-resolution Analog-to-Digital conversion, DSP algorithms in FPGA, and high-speed Digital-to-Analog conversion, the feedback controlling voltage can be obtained for each bunch. Initial tests were also conducted, and the results indicate that an ENOB better than 9.5 bit (with input frequency up to 300 MHz) and a delay adjustment precision better than 10 ps is achieved, which is good enough for the application. The frequency response of the whole system also concords well with the simulation results, and the suppression in amplitude response at the critical frequency points is better 35 dB while the uncertainty of phase response is better than 2º, all meeting the application requirement..

## Acknowledgment

The authors would like to appreciate Dr. Leng Yongbin and Dr. Lai Longwei for their kindly help. And last, the authors thank all of the SSRF collaborators who helped this paper possible.